\documentclass[preprint2]{aastex}

\newcommand{\etal}{{et~al.\null}}
\newcommand{\eg}{{e.g.,}}

\newcommand{\ie}{{i.e.,}}
\newcommand{\kms}{km~s$^{-1}$}
\newcommand{\simgt}{{\raise-.5ex\hbox{$\buildrel>\over\sim$}}}
\newcommand{\parcsec}{{\tt ''}\mskip -7.6mu.\,}
\newcommand{\parcmin}{{\tt '}\mskip -6.0mu.\,}
\newcommand{\pdeg}{^{\circ}\mskip -7.6mu.\,}
\begin{document}

\title{Planetary Nebulae in Face-On Spiral Galaxies. \\ 
I. Planetary Nebula Photometry and Distances}

\shorttitle{Planetary Nebula Photometry and Distances}
\shortauthors{Herrmann \etal}

\author{Kimberly A. Herrmann\altaffilmark{1,2}, Robin Ciardullo\altaffilmark{1,2}}
\affil{Department of Astronomy \& Astrophysics, The Pennsylvania State University \\ 525 Davey Lab, University Park, PA 16802}
\email{herrmann@astro.psu.edu, rbc@astro.psu.edu}

\author{John J. Feldmeier\altaffilmark{2}}
\affil{Department of Physics \& Astronomy, Youngstown State University, Youngstown, OH 44555-2001}
\email{jjfeldmeier@ysu.edu}

\and

\author{Matt Vinciguerra\altaffilmark{2}}
\affil{Department of Astronomy \& Astrophysics, The Pennsylvania State University \\ 525 Davey Lab, University Park, PA 16802}
\email{mattv@astro.psu.edu}

\altaffiltext{1}{Visiting Astronomer, Cerro Tololo Inter-American Observatory. CTIO is operated by AURA, Inc.\ under contract to the National Science Foundation.}

\altaffiltext{2}{Visiting Astronomer, Kitt Peak National Observatory, National Optical Astronomy Observatories, which is operated by the Association of Universities for Research in Astronomy, Inc.\ (AURA) under cooperative agreement with the National Science Foundation.  The WIYN Observatory is a joint facility of the University of Wisconsin-Madison, Indiana University, Yale University, and the National Optical Astronomy Observatories.}

\begin{abstract}
As the first step to determine disk mass-to-light ratios for normal spiral galaxies, we present the results of an imaging survey for planetary nebulae (PNe) in six nearby, face-on systems: IC~342, M74 (NGC~628), M83 (NGC~5236), M94 (NGC~4736), NGC~5068, and NGC~6946.   Using Blanco/Mosaic~II and WIYN/OPTIC, we identify 165, 153, 241, 150, 19, and 71 PN candidates, respectively, and use the Planetary Nebula Luminosity Function (PNLF) to obtain distances.   For M74 and NGC~5068, our distances of $8.6 \pm 0.3$~Mpc and $5.4^{+0.2}_{-0.4}$~Mpc are the first reliable estimates to these objects; for IC~342 ($3.5 \pm 0.3$~Mpc), M83 ($4.8 \pm 0.1$~Mpc), M94 ($4.4^{+0.1}_{-0.2}$~Mpc), and NGC~6946 ($6.1 \pm 0.6$~Mpc) our values agree well with those in the literature. In the larger systems, we find no evidence for any systematic change in the PNLF with galactic position, though we do see minor field-to-field variations in the luminosity function.  In most cases, these changes do not affect the measurement of distance, but in one case the fluctuations result in a $\sim$0.2~mag shift in the location of the PNLF cutoff.   We discuss the possible causes of these small-scale changes, including internal extinction in the host galaxies and age/metallicity changes in the underlying stellar population.
\end{abstract}

\keywords{distance scale --- galaxies: distances and redshifts --- planetary nebulae: general}

\section{INTRODUCTION}

Planetary Nebulae (PNe) are dying low-mass ($M \lesssim 8M_{\odot}$) stars composed of an expanding rarefied shell of gas surrounding a soon-to-be white dwarf \citep[\eg][]{iau209}.  The central stars of PNe can be among the brightest stellar objects in a galaxy with maximum luminosities above $\sim$6000~$L_{\odot}$ \citep{vw94}.  Due to the physics of photoionization, up to $\sim$10\% of this flux can be reprocessed into a single emission line of doubly ionized oxygen at 5007~\AA, transforming the objects into very bright monochromatic stars \citep{p1, jc99, marigo}.  Moreover, because all stars with main-sequence masses between $\sim$1~$M_{\odot}$ and $\sim$8~$M_{\odot}$ eventually evolve into PNe, these objects are plentiful in all stellar populations with ages between $10^8$ and $10^{10}$~yr.

The above properties make PNe extremely useful probes of (1) the internal kinematics of spiral \citep{M31PNS, M33, M94PNS}, elliptical \citep{PNS, deLorenzi}, and interacting \citep{M51, CenA} galaxies, (2) the chemical evolution of the Local Group \citep[\eg][]{d+97, rmc98, m+07}, (3) the stellar populations of early-type systems \citep{b+06, r06}, (4) the extragalactic distance scale \citep[\eg][]{p12, gdansk}, and (5) the dynamical evolution of galaxy clusters \citep{ipn3, arnaboldi04, gerhard07}.   We seek to use PNe as kinematic test particles in face-on spiral galaxies in order to use their motions to measure the distribution of mass in spiral disks.  (This analysis will be the subject of the second paper in this series, hereafter referred to as Paper~II.)  The first step in doing this is to identify large samples of PNe for analysis.  With these data, we can also measure the PNe's photometric properties, use their luminosity function to estimate distance, and explore how the PNLF behaves in a range of star-forming environments.

In this paper, we describe an [\ion{O}{3}] $\lambda 5007$ and H$\alpha$ survey for planetary nebulae in six nearby, face-on spiral galaxies.  In \S2 we present the details of our observing runs and image reduction.  In \S3 we explain how we identified our PN candidates and measured their magnitudes. We discuss the discrimination of possible contaminants in \S4, including how we used photometry and follow-up spectroscopy to eliminate \ion{H}{2} regions, supernova remnants, possible background galaxies, and four asteroids from the sample.  In \S5, we show how we determined the completeness limits of our samples and also compare our narrow-band selected PNe sample in M94 to a sample of planetaries found via slitless spectroscopy.  In general, we find good agreement between the two data sets, although the additional information provided by our H$\alpha$ photometry does allow us to eliminate a few compact \ion{H}{2} regions from consideration.  In \S 6, we derive the PNLF distance to each of our six galaxies and compare these new measurements to estimates in the literature.   In all cases, our values are consistent with previous distance determinations, and in three cases, our measurements are significantly more precise than those of previous studies.  In \S7 we study the shape of the PNLF, and test for differences between various PN subsamples.  We show that small variations in the shape of the PNLF are not uncommon, although these changes rarely affect our ability to measure distance.  We conclude by discussing the possible implications of these fluctuations.

\section{OBSERVATIONS AND IMAGE REDUCTION}

We chose for our study six nearby spiral galaxies, all with low recessional velocities ($v < 700$~\kms), all relatively face-on ($i \lesssim 35^\circ$), and, with one exception, all with large optical angular diameters ($r > 10\arcmin$).  These galaxies are listed in Table~\ref{Galaxies}, along with their photometric properties and metallicities.  The two southernmost objects in the sample, M83 and NGC~5068, were imaged with the Mosaic~II camera on the CTIO 4~m Blanco telescope.  This instrument is composed of 8 $2048 \times 4096$ SITe CCDs, giving it a pixel scale of $0\farcs 26$ pixel$^{-1}$ and a field-of-view large enough to encompass the entire galaxy (\ie\ $36\arcmin \times 36\arcmin$ or $\simgt$10 disk scale lengths for M83 and more for NGC~5068).  The remaining galaxies were imaged with OPTIC, an orthogonal transfer CCD camera on the WIYN 3.5~m telescope at Kitt Peak \citep{h+03, t+02}.  OPTIC, which consists of two Lincoln Lab CCID28 $2048 \times 4096$ chips separated by $14\arcsec$, produces a pixel scale of $0\farcs 14$ pixel$^{-1}$ and a field-of-view, $9\farcm 56 \times 9\farcm 56$, that is slightly smaller than the typical size of our galaxies.  As a result, our surveys with this instrument extended over $\sim$1.2, $\sim$1.8, $\sim$3.5, and $\sim$7 disk scale lengths for IC~342, NGC~6946, M74, and M94, respectively.  Although OPTIC is designed to track rapid image motion by real-time clocking of charge in both the vertical and horizontal direction, the lack of suitably bright field stars prevented us from using this feature.  Table~\ref{tabObs} presents a log of our observations.

\begin{figure}[h]
\includegraphics[scale=0.35]{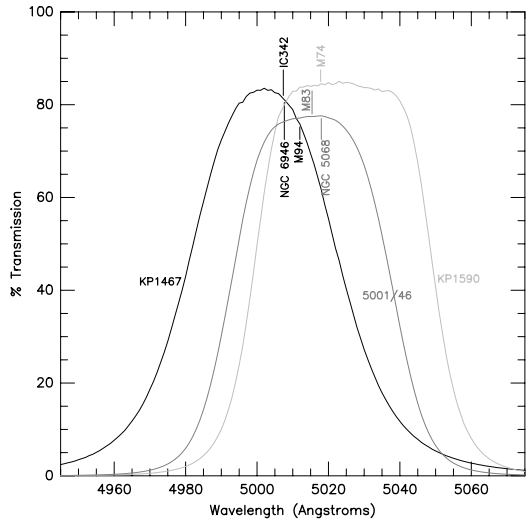}
\caption{Transmission curves for the [\ion{O}{3}] $\lambda 5007$ filters used in this survey. These curves represent the filter at the outside  temperature and in the converging beam of the telescope.  The wavelengths of [\ion{O}{3}] $\lambda 5007$ at the systemic velocity of each galaxy are marked. \label{figFilters}}
\end{figure}

For our survey, we followed the standard observing procedure for imaging extragalactic planetary nebulae \citep[\eg][]{mudville, fcj97}.  First we chose a narrow-band filter to isolate the [\ion{O}{3}] $\lambda 5007$ emission line at the redshift of the target galaxy.  For M83 and NGC~5068, this filter was custom made, with a central wavelength of $\lambda_c = 5001$ and a full-width-half-maximum (FWHM) of 46~\AA\ in the converging f/2.87 beam of the telescope.    For the remaining galaxies, standard Kitt Peak $\sim$50~\AA\ FWHM filters were used.  These filters are listed in Table~\ref{tabObs}; models for their transmission curves at dome temperature are displayed in Figure~\ref {figFilters} along with the wavelength of [\ion{O}{3}] $\lambda 5007$ shifted to the systemic velocity of each target galaxy.

Each galaxy was imaged via a series of 30~minute exposures, with dithering sequences sometimes used to fill in the gaps between chips.  For M83 and NGC~5068, corresponding 1 minute exposures through a standard $V$ filter served as off-band images; for the WIYN observations, continuum observations were obtained via a set of 7.5~min exposures through the WIYN~32 filter ($\lambda_c = 5288$, FWHM = 230~\AA\null).  In general, these off-band exposures were scaled to go at least 0.2~mag deeper than their on-band counterparts to avoid false detections at the frame limit.   To help discriminate PNe from (lower-excitation) \ion{H}{2} regions, each galaxy was also imaged in H$\alpha$.  For M83 and NGC~5068, the H$\alpha$ filter had a central wavelength of $\lambda_c = 6563$ and a FWHM of 80~\AA, while for the other four galaxies, $\lambda_c = 6555$ and FWHM = 46~\AA.  Finally, for IC~342, M74, M83, and NGC~5068, images were also taken through a standard $R$ filter.  These served as off-band images for the H$\alpha$ exposures.

All images were reduced using packages in IRAF\footnotemark[1] \citep{v98}.  The Mosaic~II exposures of M83 and NGC~5068 were reduced entirely within the {\tt mscred} package:  the data were trimmed, bias subtracted, and flatfielded via {\tt ccdproc}, placed on an astrometric system by applying {\tt mscmatch} to $\sim$2800~stars from the USNO-B1.0 catalog \citep{usno-b}, projected onto the tangent plane with {\tt mscimage}, sky subtracted with {\tt mscskysub}, aligned with {\tt mscimatch}, and co-added (with cosmic-ray removal) with {\tt mscstack}.   During this process, special care was taken to define the images' world coordinate system, as we found it necessary to run {\tt msccmatch} separately on each individual readout section of each chip (\ie\ 16 runs per exposure) to obtain a proper stacking and an astrometric precision of $\lesssim0\parcsec3$.  We found the NOAO Deep Wide-Field Survey MOSAIC Data Reductions website\footnotemark[2] to be very useful.

\footnotetext[1]{IRAF is distributed by the National Optical Astronomy Observatories, which are operated by the Association of Universities for Research in Astronomy, Inc., under cooperative agreement with the National Science Foundation.}

\footnotetext[2]{http://www.noao.edu/noao/noaodeep/ReductionOpt/frames.html}

The OPTIC images of IC~342, M74, M94, and NGC~6946 were reduced using a combination of IRAF's {\tt imred} and {\tt mscred} packages.  First, all individual exposures (including biases and flats) were converted to Multi-Extension FITS (MEF) format using {\tt mkmsc}; this enabled us to use the {\tt mscred} package to correct for the four different overscan regions and the $14\arcsec$ gap between chips.  The OPTIC reduction notes\footnotemark[3] available off the WIYN website were very useful for this first step.  Next {\tt ccdproc} was used to trim and flatfield as above.  Again, special care was needed to define the frames' world coordinate system, as each readout section (4 per frame) required a manual initialization with a centrally located star (via {\tt ccsetwcs}) and then a refinement (with {\tt ccfind} and {\tt ccmap}).  For M94, only $\sim$20 USNO-B1.0 stars \citep{usno-b} per quadrant were available for the astrometric solution; for M74, this number was closer to 40.  (Since IC~342 and NGC~6946 are both close to the Galactic plane, more than 100 stars per quadrant were available for their solutions.)  In all cases, our final astrometry attained a precision of $\lesssim0\parcsec3$.  Finally, {\tt mscimage} was run to project the frames onto the tangent plane, {\tt linmatch} was used to determine the scaling between exposures, and {\tt imcombine} was used to remove cosmic rays and stack the images.

\footnotetext[3]{http://www.wiyn.org/OPTICreduce.pdf}

\section{PLANETARY NEBULA CANDIDATE IDENTIFICATION AND PHOTOMETRY}

PN candidates were identified by examining the stacked [\ion{O}{3}] images, their H$\alpha$ and off-band counterparts, and difference images created by scaling, then subtracting the off-band images from their on-band frames.  PN candidates had to (1) have an image profile consistent with that of a point source, (2) be present in [\ion{O}{3}] but invisible in the off-band image (or extremely weak in $V$), and (3) be invisible or weak in H$\alpha$ (and $R$ when available).  This latter condition served to eliminate \ion{H}{2} regions and supernova remnants from the sample (see below).

Photometric measurements of all the candidates in [\ion{O}{3}] $\lambda 5007$ and H$\alpha$ were performed with the IRAF version of DAOPHOT \citep{s87, sdc90}.  First, the point spread function (PSF) fitting routines in {\tt allstar} were used to define the PN magnitudes on the difference images relative to those of field stars on the on-band frames.  These results were then checked by using DAOPHOT's {\tt substar} option to subtract off a scaled-PSF representation of each PN from its position on the on-band and difference frames.  Although the majority of these subtractions produced little or no residual, a few objects, usually in or near the galaxy's spiral arms, were found to be significantly under- or over-subtracted, sometimes by $\sim$0.5~mag.  The cause of this error was an incorrect estimate of the background:  in regions of active star formation, bright emission from \ion{H}{2} regions, supernova remnants, and the diffuse ISM can play havoc with the underlying ``sky,'' particularly in H$\alpha$.  When this occurred, the PN magnitudes were manually adjusted until the residuals on the star-subtracted images appeared reasonable.

Next our relative [\ion{O}{3}] $\lambda 5007$ and H$\alpha$ magnitudes were placed on the standard AB system by comparing large-aperture measurements of our field stars to similar measurements of spectrophotometric standards.  For our southern hemisphere measurements, these standards came from \citet{sb83}; the stars in the north were drawn principally from the list of \citet{s77}.  Finally, the narrow-band AB magnitudes were converted to monochromatic fluxes using models for the filter transmission curves in the converging beams of the telescopes (Figure~\ref{figFilters}), knowledge of the galaxies' systemic velocities (listed in Table~\ref{Galaxies}), and the photometric procedures for emission-line objects described by \citet{jqa87}.  In the case of our [\ion{O}{3}] measurements, these fluxes were converted to [\ion{O}{3}] $\lambda 5007$ magnitudes using
\begin{equation}
m_{5007} = -2.5\log F_{5007} - 13.74,
\label{eqflux}
\end{equation}
where $F_{5007}$ is given in ergs~cm$^{-2}$ s$^{-1}$ \citep{p1}.  At this time, we also used the photometric uncertainties derived by DAOPHOT to determine the mean photometric error of our measurements as a function of [\ion{O}{3}] $\lambda 5007$ brightness.  This photometric error function, which is given in Table~\ref{photerror}, was used to help determine the PNLF distances of \S 4.

\section{DISCRIMINATING PNE FROM CONTAMINANTS}

Planetary nebulae are not the only sources that are detectable via our on-band/off-band survey technique.  Deep images through an [\ion{O}{3}] $\lambda 5007$ filter will also detect \ion{H}{2} regions, supernova remnants, background [\ion{O}{2}] emitting galaxies (at $z \sim 0.34$), high-redshift Ly$\alpha$ sources (at $z \sim 3.12$), and even asteroids.  We discuss each of these below.

\subsection{\ion{H}{2} Regions}

At the typical distance of our galaxies ($\sim$5~Mpc), $1\arcsec$ corresponds to $\sim$25~pc.  Consequently, at groundbased resolution, compact \ion{H}{2} regions will be unresolved, and can be misclassified as planetaries.  Fortunately, a quantitative way exists to eliminate most of these contaminants using two of a PN's brightest emission lines.  \citet{p12} have shown that bright PNe populate a distinctive region in [\ion{O}{3}]-H$\alpha$ emission-line space.  Specifically, when $R = I(\lambda 5007)/I({\rm H}\alpha+$[\ion{N}{2}]) is plotted against absolute [\ion{O}{3}] magnitude, true PNe occupy a cone defined by
\begin{equation}
4 > \log R > -0.37 \, M_{5007} - 1.16.
\label{eqsquiggle1}
\end{equation}
This equation, which is valid over the top $\sim$4~mag of the PN luminosity function, is equivalent to the statement that the ratio of [\ion{O}{3}] $\lambda 5007$ to $H\alpha$ depends on the luminosity of the planetary, via
\begin{equation}
4 > R > 3.14 \left( {L \over L^*} \right)^{0.92}
\label{eqsquiggle2}
\end{equation}
where $L^* = 2.4 \times 10^{36}$~ergs~s$^{-1}$ is the absolute [\ion{O}{3}] $\lambda 5007$ luminosity of a PN at the bright-end of the PNLF.  Practically speaking, this means that PNe in the top $\sim$1.5~mag of the [\ion{O}{3}] luminosity function always have [\ion{O}{3}] $\lambda 5007$ brighter than H$\alpha$.  This contrasts with the line ratios of the vast majority of \ion{H}{2} regions, which have H$\alpha$ much brighter than [\ion{O}{3}] \citep[\eg][]{shaver, kniazev, pena}.

For three of our galaxies, our narrow-band photometry was sufficient to locate each PN candidate in the [\ion{O}{3}] $\lambda 5007$-H$\alpha$ emission line diagram.  Candidates which failed to satisfy the criterion given by equation~(\ref{eqsquiggle1}) within the uncertainties were excluded from further consideration.  Unfortunately, due to clouds and poor seeing, the H$\alpha$ images of the other three galaxies (IC~342, M74, and NGC~6946) did not go deeply enough to detect the H$\alpha$ emission from all the possible PNe.  Specifically, in IC~342 only $\sim$55\% of our [\ion{O}{3}] $\lambda 5007$ point sources were detected in H$\alpha$; in M74, the H$\alpha$ recovery fraction was just $\sim$20\%, and in NGC~6946, none of our bright PN candidates were detected in H$\alpha$.  In the case of IC~342, its nearness ($3.5 \pm 0.3$~Mpc; see \S 4) and extreme extinction ($A_V = 1.94$~mag) helped to enhance H$\alpha$ with respect to [\ion{O}{3}] $\lambda 5007$, somewhat lessening the shallow H$\alpha$ imaging problem.

For IC~342 and M74, we compensated for the lack of H$\alpha$ photometry using spectroscopy.  Approximately 68\% of the PN candidates in IC~342 have spectra obtained from the Hydra spectrograph on the WIYN telescope.  (A full description of these data is presented in Paper~II\null.)  For M74, the fraction of PNe with spectra is $\sim$65\%.  Although these data were obtained for the purpose of obtaining radial velocities and only covered the wavelength range $\sim$4500~\AA\ to $\sim$5500~\AA, it is possible to use them to estimate the H$\alpha$ flux of each candidate.

To do this, we began by subtracting the continuum from each spectrum, and measuring the relative line fluxes of [\ion{O}{3}] $\lambda 5007$, [\ion{O}{3}] $\lambda 4959$, and H$\beta$.  Appropriately, the ratio of the two oxygen lines was (on average), three-to-one, indicating that the response function within this small wavelength range was approximately flat.  We then examined the throughput of the spectrograph using observations of a standard star.  Again, the data indicated only a slight ($\sim$8\%) decrease in efficiency between the wavelengths of [\ion{O}{3}] $\lambda 5007$ and H$\beta$.  (This is close to the $\sim$5\% drop expected from the blaze function of the grating.)  Using this knowledge, we derived the true [\ion{O}{3}] $\lambda 5007$-H$\beta$ ratio for each object, then scaled these values to H$\alpha$, using an estimate of foreground Galactic extinction \citep{sfd98}, a \citet{ccm89} reddening law with $A_V = 3.1$, and an expected intrinsic H$\alpha$ to H$\beta$ ratio of 2.86 \citep{brocklehurst}.  For cases where the H$\alpha$ flux was not detected photometrically, this spectroscopic measurement was used to estimate $R$ and eliminate compact \ion{H}{2} regions from the sample.  We note that this technique may overestimate the [\ion{O}{3}]-to-H$\alpha$ flux ratio for PNe that are heavily reddened by circumstellar dust.  However, these objects are likely to fall well below the PNLF's bright-end cutoff, and not be an important part of the sample.

In the extreme case of NGC~6946, where no spectroscopy was available, the best we could do is place an upper limit on the H$\alpha$ brightness of our PN candidates.  To do this, we began by using the brightest point source magnitude returned by {\tt allstar} during our analysis of the H$\alpha$ frame.  We then scaled the frame's PSF to this magnitude, and used {\tt substar} to subtract off this ``brightest star'' from the location of each PN.  We then visually inspected the resulting image.  As expected, due to differences in the underlying background, this procedure occasionally resulted in an oversubtraction.  When this occurred, we manually adjusted the limiting {\tt allstar} magnitude until the subtraction residual showed no sign of a flux deficit.  This magnitude was then used to define the upper limit to the object's H$\alpha$ flux.  Unfortunately, even with these limits, it was impossible to discriminate PNe from \ion{H}{2} regions with sizes under $\sim$25~pc.  Thus, all the PN identifications in this galaxy must be considered tentative.

\subsection{Supernova Remnants}

Although compact supernova remnants (SNRs) are much rarer than PNe, they may present a problem at the very bright end of the PNLF\null.  This is particularly true for the more distant galaxies, where groundbased $\sim$1\arcsec\ seeing limits our spatial resolution to $\sim$25~pc.  The difficulty is well-illustrated by \citet{roth04}, who used spatially resolved spectrophotometry to show that object 276 in the list of M31 bulge PNe presented by \citet{p2} is actually a supernova remnant.  Fortunately, most SNRs have line ratios that clearly distinguish themselves from planetary nebulae.  For example, of the $\sim$60~M31 supernova remnants surveyed by \citet{galarza}, all but one has [\ion{O}{3}] $\lambda 5007$ fainter than H$\alpha$.

More specifically, we can compare the photometric properties of PNe directly with those of supernova remnants using the optical sample of M83 SNRs created by \citet{blair}.  Their catalog contains 71 supernova remnants, selected on the basis of strong [\ion{S}{2}] $\lambda 6716,6731$ relative to H$\alpha$.  Of these, only three fall within $3\farcs 5$ of one of our PN candidates, and in each case, there is a better, more distinctive SNR candidate nearby.  None of the other 68 SNRs are within $10\arcsec$ of a PN\null.  \citet{blair} also presented 18 additional [\ion{O}{3}]-bright objects, which they believe might be extremely young SNRs.  Interestingly, all of these sources are significantly fainter than the prototype of the class, the Milky Way remnant Cas~A\null.  They are, however, comparable in brightness to the well-known SMC remnant, 1E~0102$-$7219, and in theory, such objects can be confused with PNe.  Seventeen of the \citet{blair} [\ion{O}{3}]-bright sources are more than $5\arcsec$ from any of our candidates and thus not on our list.  The remaining source (O8) is coincident with our M83 PN candidate 10, which satisfies our selection criteria in every respect.  Though we targeted this object with the Hydra spectrograph, the resulting spectrum has very strong H$\alpha$ emission.  While there is virtually no H$\alpha$ emission at the location of PN 10, there is a strong H$\alpha$ emitter within $2\arcsec$.  We believe our spectrum is contaminated by this nearby H$\alpha$ emitter due to a slight misplacement of the fiber by Hydra.

In addition, there are 17 historical supernovae in our target galaxies\footnotemark[4].  An examination of our images demonstrates that none of the remnants from these sources are in our PN sample.  We did find an emission line object within $1\parcsec5$ of the location of both SNe in M74 (SN 2003gd and 2002ap) but found nothing of interest at the locations of SNe 1983N, 1968L, and 1957D in M83 and only well-resolved emission-line sources within 5\arcsec\ of the locations of SNe 1950B, 1945B, and 1923A (also in M83).  Of the nine historical SNe in NGC~6946, we found emission line objects near three (1980K, 1948B, and 1939C), some \ion{H}{2} regions and blue stars near three others (2002hh, 1969P, and 1917A), and only star-like objects around the remaining three (2008S, 2004et, and 1968D).

\footnotetext[4]{The Central Bureau for Astronomical Telegrams (CBAT) List of Supernovae website (http://cfa-www.harvard.edu/iau/lists/Supernovae.html) was used to obtain the locations of the 17 historical SNe described here.}

\subsection{Background Galaxies}

Star-forming galaxies at $z \sim 0.34$ have their [\ion{O}{2}] $\lambda\lambda 3726,3729$ doublet redshifted into the bandpass of our [\ion{O}{3}] $\lambda 5007$ filter.  Consequently, it is theoretically possible for small ($< 4$~kpc) star-bursting objects to be mistaken for planetaries.  In practice, however, such a misidentification is extremely unlikely.  At $z \sim 0.34$, our survey volume is $\sim$8500~Mpc$^3$ for the CTIO 4~m Mosaic~II data, and only $\sim$700~Mpc$^3$ for the galaxies observed with OPTIC\null.  According to the luminosity functions of \citet{hogg98}, \citet{gallego02}, and \citet{teplitz03}, this implies that our M83 and NGC~5068 fields contain less than $\sim$200~[\ion{O}{2}] galaxies above our completeness limit, and that our OPTIC images contain no more than $\sim$20~[\ion{O}{2}] emitters.  Less than 2\% of these objects will have the extreme rest frame equivalent widths ($>$60~\AA) needed to appear on our on-band frame, but be completely invisible in the continuum \citep{hogg98}.  Moreover, many of these objects will be (at least) marginally resolved.  Thus, the number of background [\ion{O}{2}] interlopers in our sample must be negligible.

In principal, Ly$\alpha$ emitting galaxies (LAEs) are another source of concern.  These objects can have extremely large ($>$100~\AA) observer-frame equivalent widths, and, in the absence of deep continuum images, can easily be mistaken for PNe.  Large numbers of such objects have been discovered in deep [\ion{O}{3}] $\lambda 5007$ surveys of Virgo's intracluster space \citep{ipn3}, and one has even been found masquerading as a PN in the M51/NGC~5195 system \citep{M51}.  Fortunately, the surveys described here do not reach the flux levels needed to sample this population.  None of the 162 LAEs found by \citet{gronwall} in their deep 0.28~deg$^2$ survey of the Extended Chandra Deep Field South would have been detected on our images of IC~342, M94, NGC~5068, and NGC~6946, and only their two brightest sources (one an AGN) would have been seen in M83.  Our survey of M74 does go deep enough to detect high-redshift Ly$\alpha$ emitters, but this galaxy was surveyed with the small field-of-view of OPTIC\null.  Consequently, the luminosity function of \citet{gronwall} predicts that only $\sim$1 of our PN candidates should be a LAE\null.  The spectroscopy of Paper~II confirms this result: Ly$\alpha$ emitters are not an important source of contamination in our survey.

\subsection{Asteroids}

Since both M74 and M83 are located at low ecliptic latitudes ($\sim$5$^\circ$ and $\sim$18$^\circ$, respectively), one might expect to find asteroids in the images of these two galaxies.  Asteroids show up as trails on the long on-band exposures since they move quickly ($\sim$0$\parcsec4$ min$^{-1}$) across the field of view and as nearly point sources on the short off-band exposures.  This makes them very easily distinguishable from PNe.  We found no asteroids on our $\sim$10$\arcmin\times10\arcmin$ M74 images but found four on our $\sim$37$\parcmin4\times36\parcmin5$ images of M83.  The asteroid trails in the stacked on-band image correspond to asteroid sky position angles and speeds of $\sim$323$^{\circ}$ at $\sim$0$\parcsec33$~min$^{-1}$, $\sim$260$^{\circ}$ at $\sim$0$\parcsec46$~min$^{-1}$, $\sim$312$^{\circ}$ at $\sim$0$\parcsec30$~min$^{-1}$, and $\sim$299$^{\circ}$ at $\sim$0$\parcsec26$~min$^{-1}$, from brightest to faintest.  A cursory check of the IAU Minor Planet Center website suggests the brightest asteroid on our images might be minor planet (16522) Tell, which moves $\sim$0$\parcsec4$ min$^{-1}$ with a PA of $325\pdeg1$ and is expected to have been $\sim$45\arcsec\ from the observed location on each of our four frames and have virtually the same motion.  However, a more detailed analysis would be necessary to verify this possibility.

\section{COMPLETENESS OF THE FINAL SAMPLES}

\begin{figure*}[h]
\centerline{\includegraphics[scale=0.8]{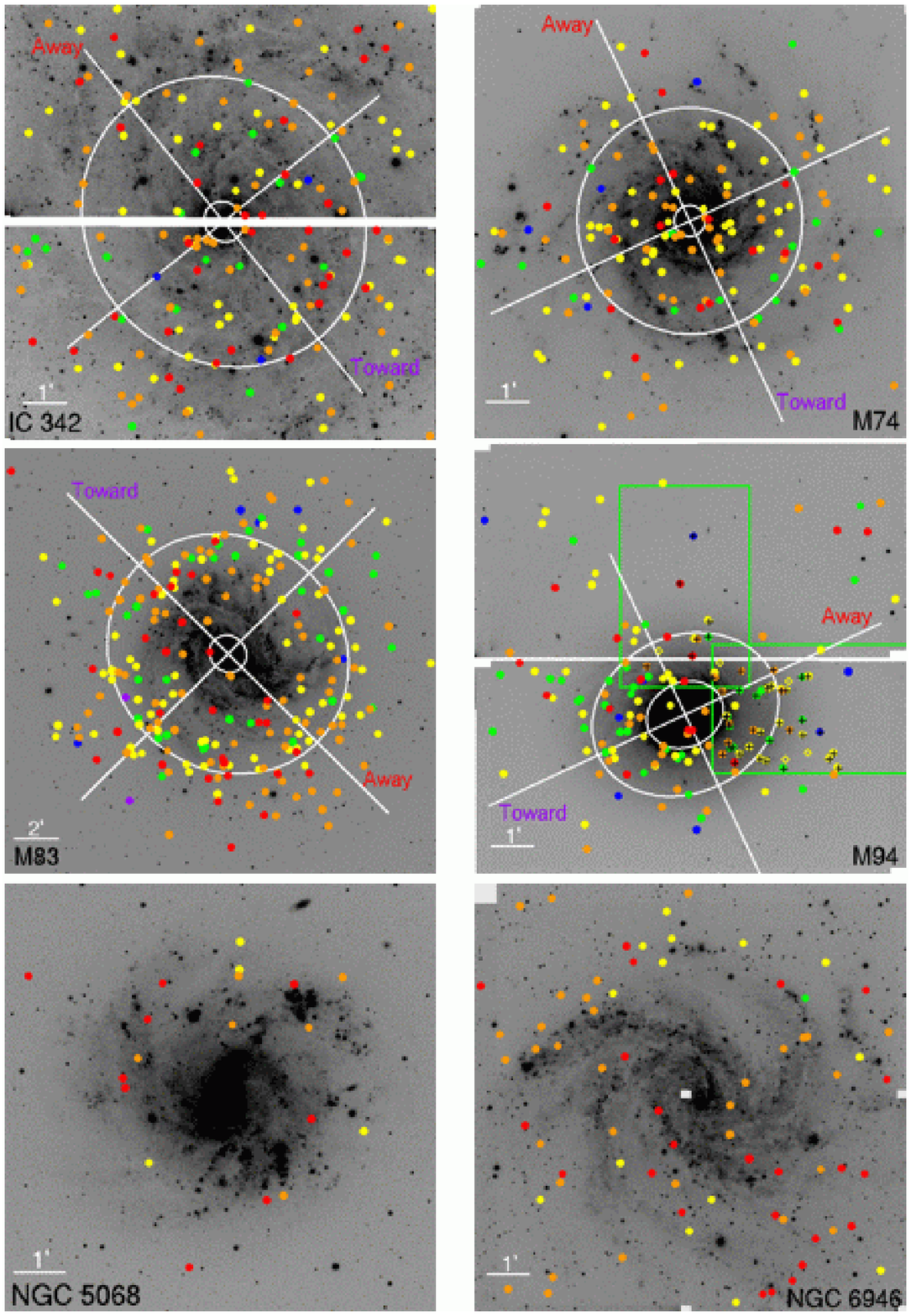}}
\caption{\small Our [\ion{O}{3}] $\lambda 5007$ images of the six galaxies studied in this survey.  North is up and east is to the left; the positions of the PN candidates are colored such that red, orange, yellow, green, and blue represent objects in the top 0.5, 1.0, 1.5, 2.0, and 2.5 mag of the PNLF, respectively.  The major and minor axes are shown as white lines.  For IC~342, M74, M83, and M94, the inner ellipse marks the  high surface-brightness region of the galaxy within which PN identifications were severely incomplete, while the outer ellipse shows the dividing line between our inner and outer samples.  For M94, the green rectangles indicate the approximate field of view of the \citet{M94PNS} slitless spectroscopy survey, the black crosses indicate candidates in both samples, and the yellow diamonds indicate their candidates that are not present in our sample.
\label{GalImages}
\notetoeditor{This figure of six images should take up a full page.} }
\end{figure*}

\begin{figure}[h]
\plotone{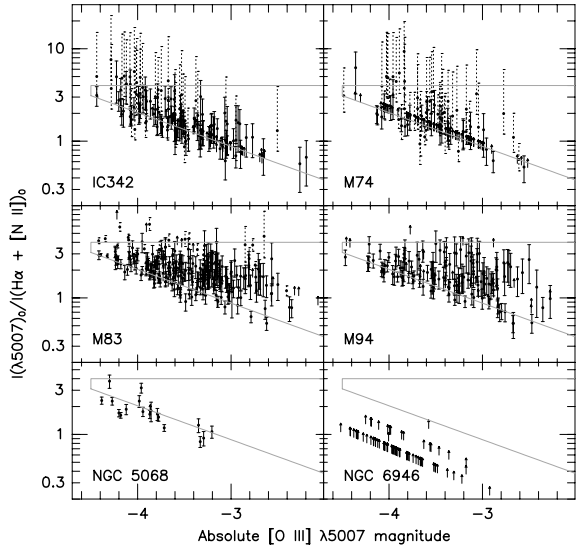}
\caption{The flux ratio between [\ion{O}{3}] $\lambda 5007$ and H$\alpha$+[\ion{N}{2}] for the planetary nebula candidates of our six galaxies.  These ratios have been corrected for foreground Galactic reddening using the extinction estimates of \citet{sfd98}, but the effects of internal and  circumstellar reddening are still present in the figure.  The dashed lines  indicate values determined from [\ion{O}{3}] $\lambda 5007$-H$\beta$ spectroscopy; arrows indicate lower limits.  The black cone-shaped outline illustrates the locus of Local Group PNe \citep{p12}.   Since the vast majority of \ion{H}{2} regions have H$\alpha$ brighter than [\ion{O}{3}] $\lambda 5007$, a diagram such as this is extremely useful for discriminating PNe from low-excitation interlopers.
\label{Frat}}
\end{figure}

After the application of the procedures of \S4, we were left with the following numbers of strong PN candidates in each galaxy: 165 in IC~342, 153 in M74, 241 in M83, 150 in M94, 19 in NGC~5068, and 71 in NGC~6946.  The positions of these objects are displayed in Figure~\ref{GalImages}; Figure~\ref{Frat} shows the location of the objects in the [\ion{O}{3}] $\lambda 5007$-H$\alpha$+[\ion{N}{2}] emission-line space.  For the latter diagram, we have plotted the photometric line ratio whenever possible; in cases of unreliable H$\alpha$ photometry, either the spectroscopic estimate or a lower limit has been displayed.   A table of PN positions, fluxes, line-ratios, and radial velocities will appear in Paper~II.  Note that it is very difficult to find PNe near the centers of galaxies and within the spiral arms, due to the bright stellar background and diffuse emission associated with regions of active star formation.  Since the ease of PN discovery increases as one moves to the lower-surface brightness outer regions of the galaxies, many of our PN candidates are at large galactocentric radii: for example, the percentages of PNe beyond 2 $V$-band scale lengths are $\sim$43\% in M74, $\sim$93\% in M83, and $\sim$73\% in M94.

The next step in our analysis was to determine our level of photometric completeness.   To do this, we performed a series of artificial star experiments using the {\tt addstar} task within DAOPHOT\null.  First we examined each image and found the locations where [\ion{O}{3}] emission from \ion{H}{2} regions, supernova remnants, and shocks made PN detections problematic.  These regions were masked out and excluded from further analysis.  At this time, we also excluded bad CCD columns, detector edges, and the regions between the two CCDs of OPTIC\null.  This removed $\sim$25\% of the disk area of M74, M83, and M94 within the galaxies' inner three disk scale lengths, and $\sim$16\% of the disk area of IC~342 and NGC~6946 within one scale radius.  (For the small galaxy NGC~5068, no exclusion regions were necessary.)  Next, we scattered a small number of artificial stars of varying [\ion{O}{3}] $\lambda 5007$ brightness throughout our on-band image.  By adding no more than $\sim$10\% of the original number of detected planetaries to each frame, we ensured that image crowding was not a factor in our analysis \citep{f95}.  Then we searched the frame for these artificial stars using the {\tt daofind} task within DAOPHOT and a threshold value of three times the background sky.  Each recovery (and non-recovery) was noted, and the process was repeated several thousand times until the completeness fraction, as a function of [\ion{O}{3}] $\lambda 5007$ magnitude and position in the galaxy, was well determined.   Finally, we fit this function to an equation of the form
\begin{equation}
f(m) = 1/2 \left[1 - \frac{\beta(m-m_{lim})}
{\sqrt{1+\beta^{2}(m-m_{lim})^{2}}}\right]
\label{eqfleming}
\end{equation}
\citep{f95}.  For purposes of this paper, we define our limiting magnitude for PN detections to be the 90\% completeness limit.  This number is listed in Table~\ref{tabObs}.

\subsection{Comparison with Previous PN Samples: M94}

One of the galaxies in our sample has previously been surveyed for planetary nebulae.  M94 was observed with a dual-beam, ``counter-dispersed imaging'' spectrograph, designed to create H$\alpha$ images with one arm, and measure PN velocities via [\ion{O}{3}] $\lambda 5007$ slitless spectroscopy with the other \citep{M94PNS}.  Although PN photometry through an instrument such as this is difficult and subject to a range of systematic errors, it is still meaningful to compare the two samples.

\citet{M94PNS} studied two regions along M94's major and minor axes, and found 53 and 14 PN candidates, respectively.  Their fields are displayed in Figure~\ref{GalImages}.  Of the 67 PN candidates found via slitless spectroscopy, 44 are located within $5\arcsec$ of one of our candidates, 16 are likely \ion{H}{2} regions and not included in our sample, one is outside our field-of-view, one appears to be a duplicate, and five are below the limit of our completeness and not present in our sample.   Conversely, 22 of the PNe found in our survey are not cataloged by \citet{M94PNS}.  Of these, eight are near the edge of the spectrograph's field, and thus may have been missed for that reason.  Of the remaining 14, two are within $\sim$0.5~mag of the bright end of the PNLF, nine are between 1 and 1.5 mag down the luminosity function, and three are extremely faint.

A comparison of the objects' [\ion{O}{3}] magnitudes shows generally good agreement.  As described above, 44 PNe were detected in both surveys.  Of these, four have slitless spectroscopy magnitudes that are more than 0.8~mag fainter than our measurements, and two have no measured magnitude at all.  However, the remaining 38 have magnitudes that are consistent with our own.  The scatter between the two data sets is $\sigma = 0.22$~mag, and the systematic zero-point offset between the two samples is only $0.019 \pm 0.036$~mag, with our values being slightly fainter.

Finally, when we examine the objects that we tagged as likely \ion{H}{2} regions, we again find good agreement.  Of the 16 sources, 10 were classified by \citet{M94PNS} as \ion{H}{2} regions based on their extremely bright [\ion{O}{3}] magnitudes.  Conversely, one object that \citet{M94PNS} considered too bright to be a PN is classified as a planetary in our sample.  This disagreement arises from the fact that \citet{M94PNS} assumed a distance of 6~Mpc in their analysis.  As we describe below, our distance to the galaxy is substantially smaller than this value.

Of the 44 candidates in common between the two samples, we have spectra of 37.  A comparison of the velocities from these two samples will be included in Paper~II.

\section{PNLF DISTANCES AND COMPARISON TO EXISTING ESTIMATES}

The Planetary Nebula Luminosity Function is an excellent standard candle:  observations in $\sim$50 elliptical, spiral, and irregular galaxies demonstrate that the bright-end of this function is well represented by an equation of the form
\begin{equation}
N(M) \propto e^{0.307M}\left(1-e^{3(M^*-M)} \right),
\label{eqPNLF}
\end{equation}
where $M^*$ is the absolute magnitude of the most luminous planetary \citep{p2, mudville}.  An external calibration using galaxies with well-determined Cepheid distances defines the nominal value of $M^*$ to be $-4.47^{+0.02}_{-0.03}$, although this cutoff appears to fade in low-metallicity objects \citep{p12}.

\begin{figure}[h]
\epsscale{0.90}
\plotone{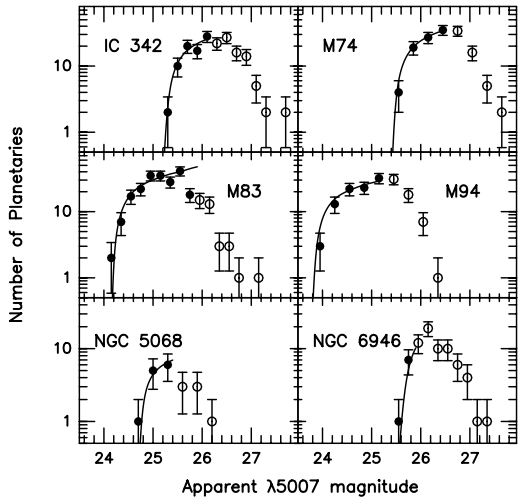}
\caption{The [\ion{O}{3}] $\lambda 5007$ PNLFs of the six galaxies analyzed in this study.  The curves represent the best-fitting empirical PNLFs  convolved with the photometric error function and shifted to the most likely distance of the galaxy.  No corrections for reddening have been applied here.  The open circles indicate points fainter than the completeness limit.
\label{figPNLFs}}
\end{figure}

To derive the PNLF distances, we began by convolving the empirical PNLF of equation~(\ref{eqPNLF}) with the photometric error functions defined in Table~\ref{photerror}.   We then fit those PNe brighter than our 90\% completeness limit to the convolved function via the method of maximum likelihood \citep{p2}.  (Since the PNLF of M83 appears to deviate from this law at faint magnitudes, we truncated its fit at $m_{5007} = 25.5$.)  Foreground Galactic extinction was taken into account by adopting the extinction estimates of \citet{sfd98} and a \citet{ccm89} reddening law, \ie\ $A_{5007} = 3.47E(B-V)$.  This analysis yielded the galaxies' distances, along with their formal fitting errors.  To these errors, we then added (in quadrature) the possible systematic uncertainties associated with our photometric standard star measurements ($\sim$0.02~mag), the filter response function ($\sim$0.02~mag), and the foreground extinction estimate ($0.16 \times A_{5007}$; see Table~\ref{tabUnc}).  Finally, a Kolmogorov-Smirnov test was used to determine whether the best-fit empirical function was indeed a valid representation of the observed data.  In all cases, this is confirmed to be true:  the worst fit is for M74, which has a strange dearth of PNe between $\sim$0.2 and $\sim$0.4~mag below the PNLF cutoff (see \S7).  However, even for this galaxy, the null hypothesis can only be excluded at the $\sim$90\% confidence level.  Figure~\ref{figPNLFs} displays the observed luminosity functions for each galaxy, along with the best fit models.

We note that our analysis of the PNLF ignores the effects of internal extinction.  This omission is justified for several reasons.  First, \citet{giovanelli} have found that Sc galaxies have very little dust extinction beyond two or three scale lengths of the galactic center.  As we mentioned above, many of our PN candidates in M74, M83, and M94 have galactocentric radii larger than this limit.  Moreover, \citet{fcj97} modeled the effects of internal extinction on the PNLF by assuming that the vertical scale heights of dust and PNe in other spiral galaxies are similar to those observed in the Milky Way.  They found that the larger scale height of the PNe make the PNLF method relatively insensitive to internal extinction.  Specifically, derived PNLF distances are never more than $\sim$0.1~mag less than their nominal value, even when the total amount of $A_{5007}$ extinction is large.  In theory, this question can be addressed directly, via spectroscopy of the PNe's H$\alpha$ and H$\beta$ emission lines, though the faintness of H$\beta$ and the effects of atmospheric dispersion make the observations challenging.

\begin{figure}[h]
\epsscale{1.0}
\plotone{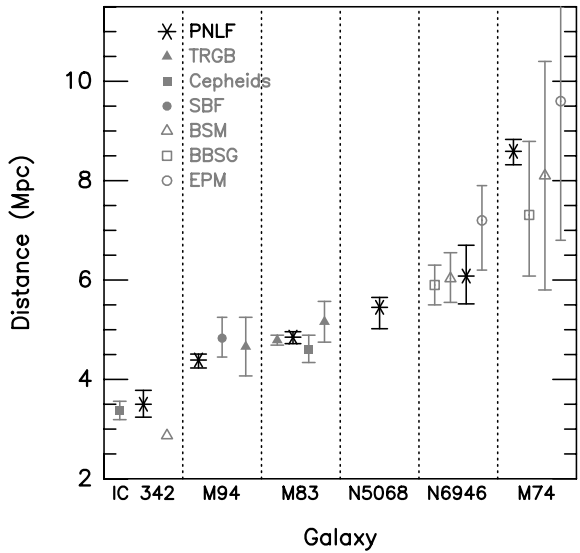}
\caption{Final PNLF distances compared to existing distances in the literature.  In most cases, only the distances with the smallest uncertainties are shown, but a few measurements with large uncertainties and one without quoted errors are provided for comparison.  See Table 5 for more information on the distances, including the reference for each value.
\label{Ds}}
\end{figure}

In the sections below, we compare our PNLF distances to values determined from the analyses of Cepheids, the Tip of the Red Giant Branch (TRGB), Surface Brightness Fluctuations (SBF), the Brightest Blue Super Giants (BBSG), the Brightest (Red and Blue) Supergiants (BSM), and the Expanding Photospheres of Supernovae (EPM).  Though some of the galaxies also have Tully-Fisher measurements, the fact that we selected our targets to have low inclination makes these determinations unreliable.   Figure~\ref{Ds} and Table~\ref{tabDMod} compare our distance moduli to measurements in the literature.  Note that the zero point for all our PNLF distances is tied to that of Cepheids via \citet{p12}.  Thus, following \citet{keyfinal}, our distance scale assumes an LMC distance modulus of $(m-M)_0 = 18.50$.

\subsection{IC~342}

IC~342, properly dubbed the Hidden Galaxy, would be one of the brightest galaxies in the sky except for its location near the Galactic plane ($b^{II} \sim 11^\circ$) and the large amount of foreground extinction ($E(B-V) = 0.558$).  The galaxy is part of the Maffei group, a system that garnered some attention due to the suggestion of a past gravitational encounter with the Local Group \citep{zvb91, v+93, peebles}.  More recently, multiple studies have presented evidence against this possibility \citep{ktg95, k+03a, f+07}.  Although the group of $\sim$16 galaxies was originally thought to have contained only two large galaxies (IC~342 and the giant elliptical Maffei~1), a recent analysis by \citet{f+07} has resulted in the addition of another member, the heavily extinguished ($A_V \sim$5.6~mag) spiral Maffei~2.

A number of different distance indicators have been used for IC~342.  \citet{sch02} used two-passband observations of 20 Cepheids to estimate a distance modulus of $(m-M)_0 = 27.58 \pm 0.18$ and a total reddening of  $A_V = 2.01$.  With an updated period-luminosity relation and metallicity measurement, this number increases slightly to $(m-M)_0 = 27.64 \pm 0.12$ \citep{f+07}.  For comparison, the magnitudes of the three brightest blue and red supergiants gives a distance modulus of $(m-M)_0 \sim 27.29$ \citep[][revised by \citet{k+97}]{kt93}.

Out of the 165 PNe found in IC~342, the 65 in the top $\sim$0.9~mag of the PNLF qualify as our complete sample.  Our application of the maximum likelihood method then gives a true distance modulus of $(m-M)_0 = 27.72 \pm 0.17$ ($3.50^{+0.28}_{-0.26}$~Mpc).  Note that the error in our distance is completely dominated by the uncertainty in the foreground extinction.  (Without this term, our error would only be $\sim$0.05~mag!)   Since the extinction derived by \citet{f+07} is completely consistent with that determined from the COBE satellite \citep{sfd98}, we adopt their $\sim$0.16~mag uncertainty in our calculation as well.

Figure~\ref{Ds} compares our distance estimate to that of the Cepheids and supergiant stars.  Our value is statistically identical to the former value, but substantially larger than that associated with the galaxy's brightest stars.

\subsection{M74}

The Perfect Spiral, M74 (NGC~628), is perhaps the most face-on galaxy in the nearby universe, with $i \sim 6.5^{\circ}$ \citep{kb92}.  This grand-design system dominates a small group which includes three dwarf irregular objects \citep{skt96}, and plays host to both our nearest
Type~Ic ``hypernova'' \citep{mazzali02} and an extremely variable ultraluminous x-ray source \citep{krauss05}.  

Though M74 is well-studied, it lacks a reliable distance estimate.  Values in the literature range from $\sim$7 to $\sim$20~Mpc \citep{sd96} and this uncertainty propagates into all aspects of its analysis, including luminosity estimates for its supernovae and x-ray sources.  M74's three least uncertain distances come from the photometry of its three brightest blue supergiants \citep[$(m-M)_0 = 29.32 \pm 0.40$;][]{skt96}, the photometry of its brightest supergiants of all colors \citep[$(m-M)_0 = 29.5^{+0.5}_{-0.7}$;][]{h+05}, and the expansion velocity-absolute magnitude relation of the Type~II-P supernova SN~2003gd \citep[$(m-M)_0 = 29.9^{+0.6}_{-0.8}$;][]{h+05}.

Our complete sample of planetaries consists of 82 objects out of a total population of 153.  By fitting the top $\sim$1.1~mag of the luminosity function, we obtain a distance modulus of $29.67^{+0.06}_{-0.07}$, or $8.59^{+0.24}_{-0.27}$ Mpc.  Though this distance differs by only $\sim$10\% from the values obtained from the Type~II-P supernova and the galaxy's supergiants, it has considerably less uncertainty.

\subsection{M83 and NGC~5068}

M83 (NGC~5236), the barred Southern Pinwheel, is one of the two central galaxies of the nearby Cen~A (NGC~5128) group, an $\sim$87 galaxy complex which also includes NGC~5068. Though often viewed as a single system, \citet{k+02, k+07} have recently shown that the group actually contains two spatially  separated associations, one centered on the interacting elliptical NGC~5128, and the other clustered about M83.  M83 itself is one of the great supernova factories in the sky, with six recorded events in the last century.

To date, the two most accurate distances to M83 have come from Cepheids and stars at the tip of the red giant branch.  The former method, based on $V$ and  $I$-band observations of 12 Cepheids with the {\sl VLT,} gives a distance  modulus of $(m-M)_0 = 28.32 \pm 0.13$ \citep{t+03, s+06}; the latter,  which comes from {\sl HST\/} F606W and F814W photometry of halo stars, yields $(m-M)_0 = 28.56 \pm 0.18$ \citep{k+07}.  A group distance,  determined via TRGB measurements to 10 of the galaxy's surrounding dwarfs, is $(m-M)_0 = 28.40^{+0.04}_{-0.05}$.

Our PN survey with the CTIO 4-m telescope detected 241~PNe in M83; 164 of these fall in the top $\sim$1.3~mag of the PNLF and comprise the complete sample before the luminosity function turns over.  Our maximum-likelihood fit to the data yields a distance modulus of $(m-M)_0 = 28.43^{+0.05}_{-0.06}$, or $4.85^{+0.11}_{-0.13}$ Mpc.  Interestingly, this value lies between the Cepheid and TRGB distances to  the galaxy, and is closer to the mean group distance than either of the two direct measurements.

For NGC~5068, 11 of our 19 PNe are above our completeness limit and in  the statistical sample.  Formally, our fit to these data produces a distance modulus of $(m - M)_0 = 28.85^{+0.09}_{-0.16}$,  a value that places the galaxy in back of the main Cen~A/M83 complex. However, \citet{p12} have noted that for systems more metal-poor than the LMC, the PNLF cutoff fades, leading to systematically larger distances.  To correct for this effect, \citet{p12} suggest using the theoretical relation derived by \citet{djv92} for the behavior of $M^*$ with metallicity, \ie
\begin{equation}
\Delta M^* = 0.928{\rm [O/H]}^2 + 0.225{\rm [O/H]} + 0.014,
\label{eqMetallicity}
\end{equation}
where the solar abundance of oxygen is assumed to be  $12 + \log {\rm (O/H)} = 8.87$ \citep{gns96}.  With this relation and the observed low oxygen abundance of the galaxy, $12 + \log {\rm [O/H]} \sim 8.32$ \citep{pvc04}, NGC~5068 moves closer to M83, to  $5.45^{+0.20}_{-0.43}$~Mpc.  This is the first direct distance measurement to this object.

\subsection{M94}

One of the many galaxies in the Canes Venatici~I cloud, M94 (NGC~4736) is surprisingly isolated with no visible companions down to a $B$-band surface brightness limit of 25~mag~arcsec$^{-2}$ \citep{kk98} within $\sim$3 degrees or 230 kpc.  This Sab spiral has a small, but bright bulge, a Low Ionization Nuclear Emission Region \citep{heckman}, and an extremely bright \ion{H}{2} ring encircling its nucleus at a distance of $\sim$1.1~kpc \citep{munoz04}.

The distance to M94 is not well established.  Although \citet{k+03b} obtained a value of $(m-M)_0 = 28.34 \pm 0.29$ from {\sl HST\/} images of the Tip of the Red Giant Branch, and \citet{t+01} derived $(m-M)_0 = 28.58 \pm 0.18$ from Surface Brightness Fluctuations of the galaxy's bulge, a significantly larger distance ($(m-M)_0 = 28.89$), based simply on the object's Hubble Flow and $H_0 = 55$~km~s$^{-1}$~Mpc$^{-1}$ is often used in the literature.

We identified 150~PNe in M94, with 71 in a complete sample which extends $\sim$1.3~mag down the PNLF\null.  Our distance modulus from  fitting these data is $(m-M)_0 = 28.21^{+0.06}_{-0.08}$, or $4.39^{+0.12}_{-0.16}$~Mpc.  This easily fits within the uncertainty of the TRGB measurement but is inconsistent with the more distant SBF value.  However, \citet{p12} have shown that the zero point of the SBF distance scale as described by \citet{t+01} is systematically larger than that of the PNLF by $0.30 \pm 0.05$~mag.  (Half of this discrepancy has since been resolved by \citet{jensen03} via a scale adjustment to the SBF calibrators.)  If this correction is taken into account, then the SBF distance becomes statistically indistinguishable from our value.

\subsection{NGC~6946}

NGC~6946, the Fireworks Galaxy, has a nuclear starburst, significant star formation activity throughout its spiral arms, and an impressive number  of recent supernovae (9 in all).   It is relatively isolated, having only  $\sim$10~low surface brightness objects in its group \citep{ksh00}, and it is located close to the Galactic plane, ($b^{II} \sim 12^\circ$). It therefore falls behind a relatively large amount of Galactic extinction \citep[$A_B \sim 1.475$;][]{sfd98}.

There are few reliable distances to NGC~6946.  \citet{skt97} used  ground-based $B$ and $V$ images to identify three of the galaxy's  brightest supergiants, and estimated a distance modulus of $(m-M)_0 =  28.90 \pm 0.18$.  Similarly, \citet{ksh00} analyzed the brightnesses of blue supergiants in the galaxy's surrounding dwarfs and inferred $(m-M)_0 = 28.85 \pm 0.15$.  Both these values imply a distance that is smaller than that found by \citet{schmidt92} by applying the Expanding Photosphere Method to the Type~II-L supernova 1980K ($(m-M)_0 = 29.29^{+0.20}_{-0.32}$).

Because the spiral arms of NGC~6946 are filled with bright \ion{H}{2} regions, identifying PNe in this galaxy is particularly difficult.  As a result, our PN sample consists of only 71 candidates, and only 11 of these are brighter than our 90\% completeness limit.  Moreover, as detailed in \S 4.1, our H$\alpha$ observations did not extend deep enough to confirm the high-excitation nature of these objects, and no follow-up spectroscopy was available in lieu of this limitation.  Still, the observed shape of our PNLF is consistent with that expected from the empirical model, and our fit of the top $\sim$0.3~mag of this function yields $(m-M)_0 = 28.92 \pm 0.21$ ($6.08^{+0.62}_{-0.56}$~Mpc).   We note that despite the limited number of PNe in the fit, our distance uncertainty is still dominated by the error associated with the galaxy's foreground extinction, $\sim$0.2~mag.  We also note that, although some of our PN candidates may in fact be \ion{H}{2} regions, it is likely that our PNLF distance will remain largely unaffected.  Specifically, a careful examination of our images suggests that four of our final 71 PN candidates might be slightly resolved.  If we eliminate these objects from our sample, our derived distance to NGC~6946 increases by less than 2\%.

\section{VARIATIONS IN THE PNLF}

The PNLF distance method has been tested as well as any general purpose extragalactic standard candle.  Despite these tests, a few lingering doubts remain.  In the {\sl HST Key Project's\/} calibration of the  Population~II distance scale, \citet{ferrarese} chose not to include the  PNLF results in their final distance determination, since its distances were systematically shorter than those found from Surface Brightness  Fluctuations.  More recently, \citet{sgm06} claimed the detection of two  kinematically distinct PN populations in the E6 elliptical galaxy NGC~4697,  with each following its own luminosity function.  Moreover, surveys of Coma and Virgo have presented evidence for the possible existence of  overluminous planetary nebulae in intergalactic space \citep{arnaboldi07,  arnaboldi08}.

The best way to test the robustness of the PNLF method is to examine subsets of PNe within a single galaxy.  Population gradients in  spiral and elliptical galaxies are common \citep{zkh94, p+90}, so if  the PNLF is observed to be constant across the face of a galaxy, that presents strong evidence that neither age nor metallicity is affecting  the luminosity function.  Unfortunately, to date, only a few PN  populations have been studied in this way.  Analyses of the PNe across M31 \citep{M31PNS}, M33 \citep{magrini00, M33}, M81 \citep{m+01}, M104 \citep{M104}, and NGC~5128 \citep{N5128} have found no radial change in the PNLF, while the gradients found in M87 \citep{c+98} and NGC~4526 \citep{fjp07} have been attributed to the presence of foreground intracluster stars. 

The planetary nebula samples of IC~342, M74, M83, and M94 are large enough for us to examine the systematics of subsamples.  We therefore divided the PNe in these galaxies in three different ways:  radially (\ie\ by defining an  inner and an outer sample), by the direction of rotation (assuming that all  PNe are rotating along with the galactic disk), and by the orientation of the galaxy (\ie\ near side versus far side).   The first of these divisions   should be sensitive to population changes within the program galaxy, the  second to wavelength-dependent variations in the transmission of the  narrow-band filter, and the third to host galaxy internal extinction. Figure~\ref{GalImages} illustrates these divisions.

\begin{figure*}[h]
\epsscale{2.0}
\plotone{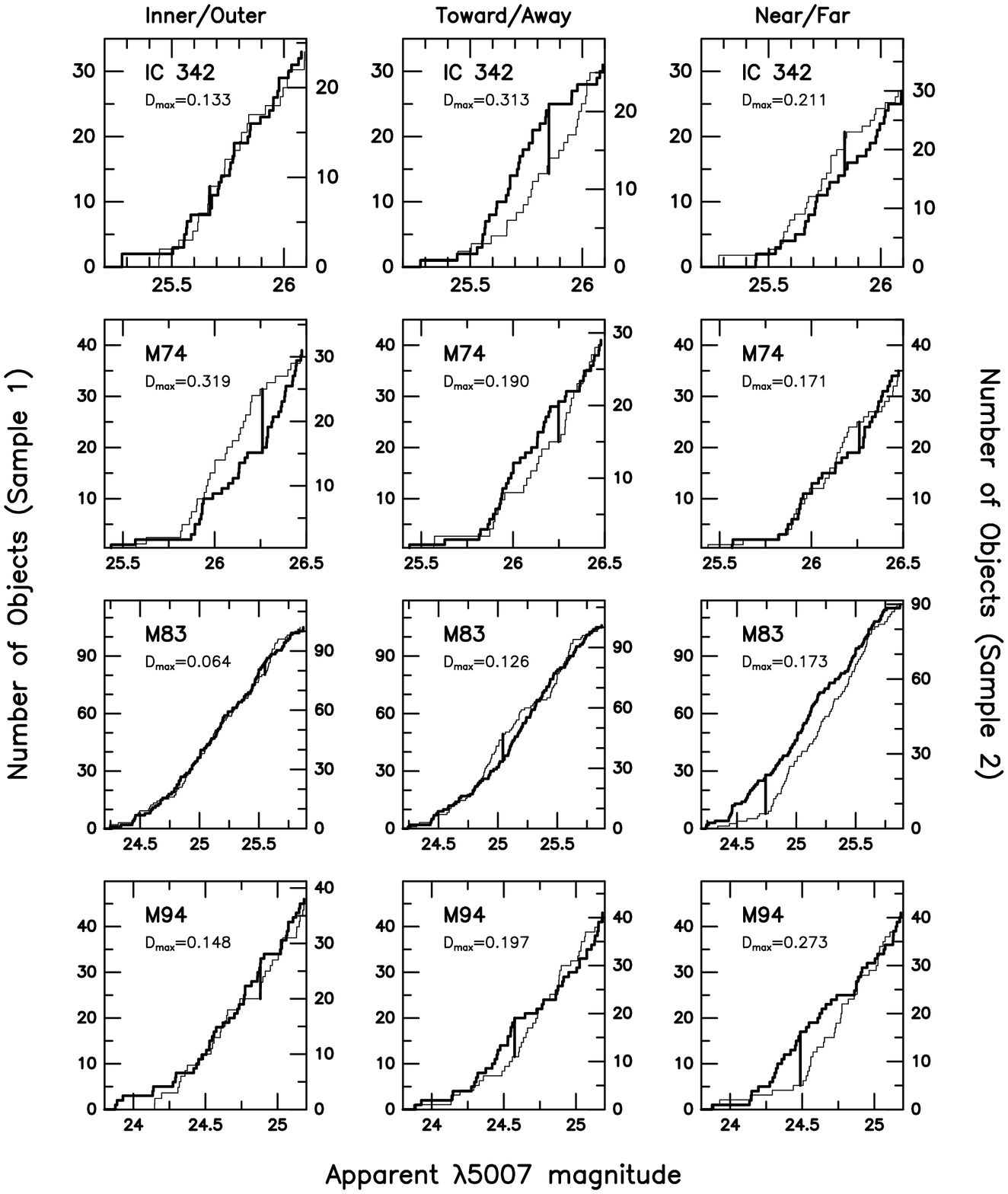}
\caption{The cumulative [\ion{O}{3}] $\lambda 5007$ luminosity functions for subsamples of PNe within four galaxies, with the Kolmogorov-Smirnov test statistic marked.  In four out of the 12 comparisons, this  statistic excludes the null hypothesis with more than 90\% confidence.   The approaching PNe of IC~342, the inner PNe of M74, and the southwest  PNe in M94 all deviate from the empirical law with more than 90\% confidence. Although the near-side and far-side samples of PNe in M83 both follow the empirical law, the cutoff magnitude changes by $\sim$0.18~mag.
\label{ksplot}}
\end{figure*}

\begin{figure}[h]
\epsscale{1.1}
\plotone{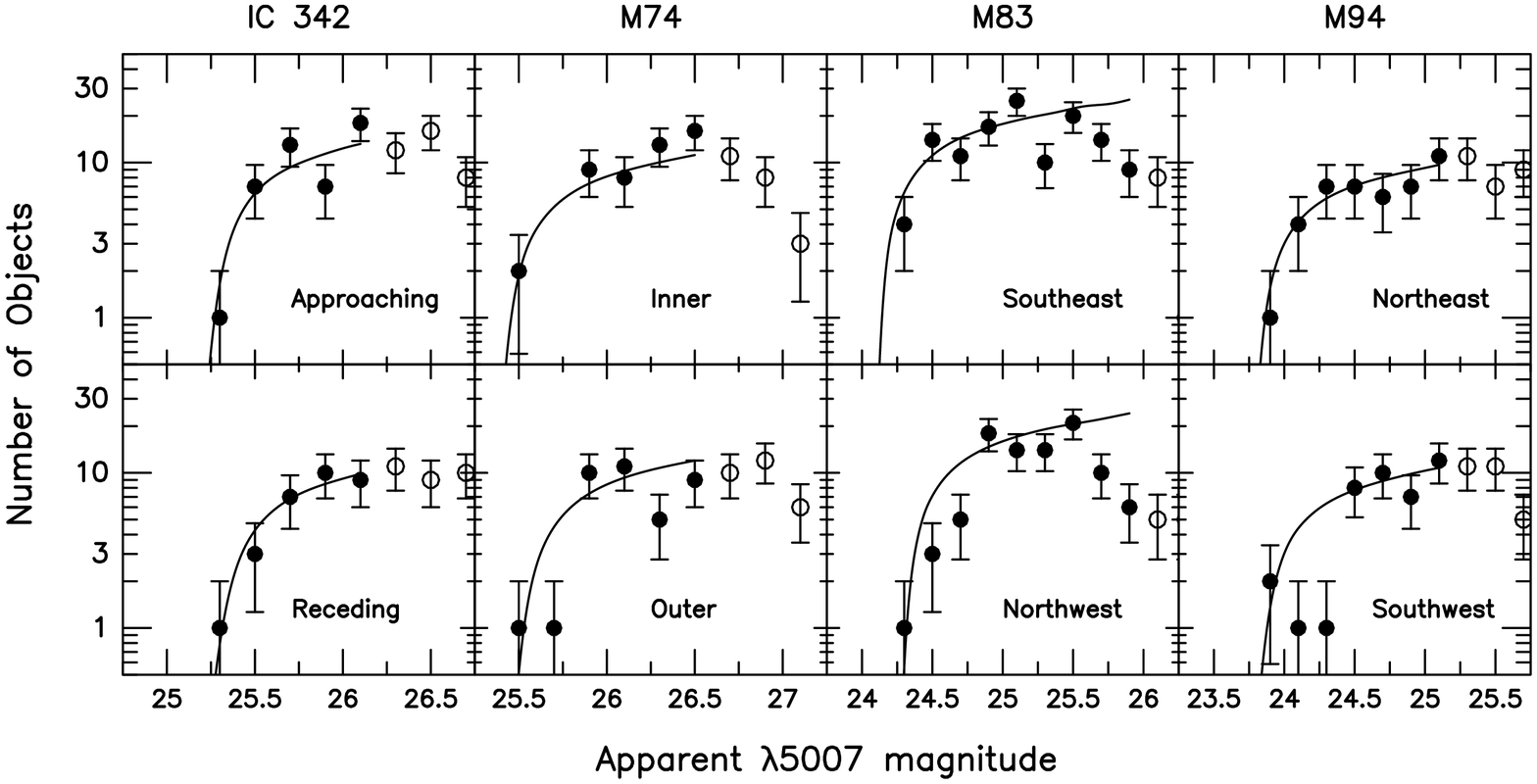}
\caption{The differential [\ion{O}{3}] $\lambda 5007$ luminosity function for the four systems where the Kolmogorov-Smirnov statistic indicates a change in the PNLF across the galaxy.   In the case of IC~342, M74, and M94, the most-likely distance modulus is not affected by these fluctuations.  In M83 however, the best-fit value for $m^*$ changes by $\sim$0.18~mag.
\label{comparePNLF}}
\end{figure}

Figure~\ref{ksplot} compares the cumulative luminosity functions of our various subsamples via a Kolmogorov-Smirnov test.  In general, there is  excellent consistency within the galaxies.  However, as the statistic  shows, a few anomalies do exist.  In IC~342, the PNLF derived from  the sample of planetaries on the approaching side of the galaxy differs  from that derived from PNe on the receding side at the 90\% confidence  level.  Interestingly, as the differential plot of Figure~\ref{comparePNLF} illustrates, both sets of objects yield identical values for the PNLF distance.  However, while the planetaries on the receding  side obey the luminosity function given in equation~\ref{eqPNLF}, the approaching PNe deviate from this law with 90\% confidence.  The PNe of M74 show a similar effect, except this time the gradient lies in  the radial direction.  Again, the PNLF distances inferred from both  samples are nearly identical, but the detailed shapes of the two  functions differ with 95\% confidence.  In this case, the problem can be traced to a deficit of objects $\sim$0.3~mag down from $m^*$. Finally, in M83 and M94, the PNe on the near side of the galaxy follow  a slightly different luminosity function than those on the far side  (at $\sim$90\% confidence).  In M94, this difference does not affect  the determination of $m^*$, since it is primarily caused by a deficit  of objects $\sim$0.5~mag below the cutoff in the sample of planetaries  north of the galaxy's major axis.  However, in M83, $m^*$ is affected:   although both samples of planetaries are well fit by equation~(\ref{eqPNLF}), the PNLF for objects on the northwest side of the galaxy yields an apparent distance modulus that is $\sim$0.18~mag farther than that  derived from the southeast sample.  This discrepancy is $\sim$3 times the formal error of the fits.

There are two possible explanations for the inconsistencies seen in the figure.  The first is internal extinction in the host galaxies. In the Milky Way, the scale height of PNe is only slightly larger than that of the dust \citep{pm87, zp91, phillips01}.  Such an intermingling can depress the magnitudes of some objects, but not others, and create a luminosity function very similar to that seen in the inner  regions of M74 or the receding side of IC~342.  Of course, to explain all the variations seen in Figure~\ref{ksplot}, one needs the extinction to be patchy, or at least have some azimuthal dependence.  Moreover, it is  well known that the amount of dust in a galaxy declines with radius \citep{giovanelli, kylafis}, so that if internal extinction is truly responsible for these effects, one would expect to see the PNLF change more with radius than with position angle.

An alternative explanation involves assuming that the PNLF's shape is population dependent, and that the differences seen in Figure~\ref{ksplot} are due to subtle changes in the age or metallicity distribution function of the progenitor stars.  To some extent, such changes should not be  surprising.  We know that the faint end of the PNLF is population dependent:  while old systems such as M31's bulge have exponentially increasing PNLFs,  star-forming systems such as the Small Magellanic Cloud exhibit a ``dip'' in  the luminosity function starting $\sim$2~mag below the cutoff  \citep{p12, jd02}.  In fact, Figure~\ref{figPNLFs} suggests that the same  dip seen in M33 and the SMC may be present in M83 as well.    However, there has never before been any evidence for small-scale features  that distort at the bright-end of the PNLF\null.  

It is becoming increasingly clear that the term ``planetary nebula'' refers  to a heterogeneous collection of objects.  PNe from single stars will have  luminosities which depend on the mass of their central star \citep{vw94}.  Since central star mass depends on progenitor mass through the initial  mass-final-mass relation \citep{weidemann, kwilliams}, populations of different ages will produce different PNLFs.  Binary stars also produce planetaries via common envelope interactions, and their contribution to the PNLF may be very different from that of single stars \citep{soker, moe06}.  Finally, \citet{pn-bs} have proposed that  the descendants of blue stragglers may contribute greatly to the  bright-end of the planetary nebula luminosity function.   The observed shape of a population's PNLF may involve a complex mixture of contributions from all these sources.  The resiliency of the PNLF technique may  simply be due to the processes which place a hard upper limit on the  emitted flux at [\ion{O}{3}] $\lambda 5007$. 

\section{DISCUSSION}

The results of our analysis demonstrate that the PNLF is still an excellent standard candle.  The distances presented to the six galaxies discussed in this paper are in good agreement with previous measurements, and the subsample comparisons performed in \S 7 indicate that there are no major population gradients in the PNLF cutoff.  However, minor unexplained variations in the luminosity function do exist, and in some cases, this can introduce an additional $\sim$0.2~mag error into the distances.  At this time, it is impossible to determine whether these variations are intrinsic to the stellar population, or caused by internal extinction in the host galaxies.  The fact that these are star-forming galaxies with large amounts of dust suggests that extinction is the cause, but the data from the \citet{sgm06} survey of the elliptical galaxy NGC~4697 and the observation of Virgo's intracluster  stars \citep{arnaboldi08} suggest otherwise.  In theory, one can discriminate between these (non-exclusive) hypotheses via spectrophotometry of the objects' H$\alpha$/H$\beta$ ratio.  In practice, however, such a study is extremely difficult.   In a typical bright planetary, H$\beta$ is $\sim$10~times weaker than [\ion{O}{3}] $\lambda 5007$, so obtaining accurate measurements is a challenge.  Moreover, PNe are subject to circumstellar extinction, as well as Galactic extinction.  Disentangling the two components will require the analysis of large samples of objects, in many different galactic environments.

\acknowledgments

We would like to thank the KPNO and CTIO personnel for friendly travel, telescope, and instrumental support and specifically acknowledge Heidi  Schweiker for measuring the transmission curves of our filters.  We would also like to thank the anonymous referee for useful comments.  This research has made extensive use of the USNOFS Image and Catalogue Archive  operated by the United States Naval Observatory, Flagstaff Station ({\tt http://www.nofs.navy.mil/data/fchpix/}), NASA's Astrophysics Data  System, and the NASA/IPAC Extragalactic Database (NED) which is operated  by the Jet Propulsion Laboratory, California Institute of Technology,  under contract with the National Aeronautics and Space Administration. This work was supported by NSF grant AST 06-07416 and a Pennsylvania Space  Grant Fellowship.

Facilities: \facility{Blanco:Mosaic~II}, \facility{WIYN:OPTIC}.


\begin{deluxetable}{llccccccc}
\tablecaption{Basic Galaxy Properties}
\tablewidth{0pt}
\tablehead{
\colhead{Galaxy}                &\colhead{Type$^a$}      &\colhead{V$_{helio}^a$} 
&\colhead{$B_T^{0a}$} &\colhead{$12 + \log {\rm O/H}^b$}  &\colhead{Size$^a$}
&\colhead{$E(B-V)^a$}             &\colhead{$i^c$}       &\colhead{PA$^c$} \\
 & & (km~s$^{-1}$) & & & & & (deg) & (deg)
}
\startdata
IC~342 &Scd  &31 &5.58 &8.85 &$21 \farcm 4$ &0.558 &$25$  &$39$ \\
M74    &Sc  &657 &9.76 &8.68 &$10 \farcm 5$ &0.070 &$6.5$ &$25$ \\
M83    &SBc &513 &8.79 &8.79 &$12 \farcm 9$ &0.066 &$24$  &$46$ \\
M94    &Sab &308 &8.75 &8.60 &$11 \farcm 1$ &0.018 &$35$  &$115$ \\
NGC 5068 &SBd &668 &10.09 &8.32 &$7 \farcm 2$ &0.102 &$26$ &$\sim$60 \\
NGC 6946 &Scd &48 &7.78 &8.70 &$11 \farcm 5$ &0.342 &$38$  &$240$ \\
\enddata
\label{Galaxies}
\tablerefs{a: NASA/IPAC Extragalactic Database (NED); b: Pilyugin \etal\ 2004; c: various sources (see Paper~II)}
\end{deluxetable} 

\begin{deluxetable}{llcccccc}
\tablecaption{Observing Log}
\tablewidth{0pt}
\tablehead{
&\multicolumn{2}{c}{[\ion{O}{3}] Filter} 
&&\multicolumn{2}{c}{Exp Time (min)} &&\colhead{Limiting} \\
\colhead{Galaxy} &\colhead{Name} &\colhead{Bandpass} &\colhead{Date}
&\colhead{[\ion{O}{3}]} &\colhead{H$\alpha$} &\colhead{Seeing} 
&\colhead{$m_{5007}$}
}
\startdata 
M83      &\dots  &5001/46 &2004 May 28-30  &180 &180 &$1\farcs 26$  &25.9 \\
NGC~5068 &\dots  &5001/46 &2004 May 28-30  &120 &150 &$1\farcs 22$  &25.4 \\
M94      &KP1467 &4996/45 &2005 May 14-16  &150 &75  &$1\farcs 00$  &25.1 \\
NGC~6946 &KP1467 &4996/45 &2005 May 14-16  &150 &30  &$0\farcs 97$  &25.9 \\
IC~342   &KP1467 &4996/45 &2005 Nov 5-6    &210 &30  &$0\farcs 79$  &26.1 \\
M74      &KP1590 &5020/51 &2005 Nov 5-6    &360 &90  &$0\farcs 83$  &26.5 \\
\enddata
\label{tabObs}
\end{deluxetable}


\begin{deluxetable}{ccccccc}
\tablecaption{Photometric Error versus Magnitude}
\tablewidth{0pt}
\tablehead{
&\multicolumn{6}{c}{Photometric Uncertainty} \\
\colhead{Magnitude} &\colhead{IC 342} &\colhead{M74} &\colhead{M83}
&\colhead{M94} &\colhead{NGC 5068} &\colhead{NGC 6946} 
}
\startdata
24.0 & \dots & \dots & \dots & 0.040 & \dots & \dots \\
24.2 & \dots & \dots & 0.027 & 0.043 & \dots & \dots \\
24.4 & \dots & \dots & 0.031 & 0.051 & \dots & \dots \\
24.6 & \dots & \dots & 0.037 & 0.064 & \dots & \dots \\
24.8 & \dots & \dots & 0.046 & 0.076 & 0.039 & \dots \\
25.0 & \dots & \dots & 0.055 & 0.092 & 0.045 & \dots \\
25.2 & 0.049 & \dots & 0.064 & 0.108 & 0.050 & \dots \\
25.4 & 0.056 & \dots & 0.072 & 0.127 & 0.057 & \dots \\
25.6 & 0.066 & 0.045 & 0.083 & 0.149 & 0.063 & 0.078 \\
25.8 & 0.080 & 0.050 & 0.095 & 0.175 & 0.074 & 0.093 \\
26.0 & 0.097 & 0.058 & 0.114 & 0.210 & 0.087 & 0.114 \\
26.2 & 0.115 & 0.066 & 0.138 & 0.254 & 0.104 & 0.134 \\
26.4 & 0.135 & 0.076 & 0.163 & 0.310 & \dots & 0.155 \\
26.6 & 0.155 & 0.093 & 0.184 & \dots & \dots & 0.180 \\
26.8 & 0.175 & 0.110 & 0.205 & \dots & \dots & 0.207 \\
27.0 & 0.198 & 0.135 & 0.229 & \dots & \dots & 0.232 \\
27.2 & 0.225 & 0.163 & 0.255 & \dots & \dots & 0.269 \\
27.4 & 0.260 & 0.192 & \dots & \dots & \dots & 0.325 \\
27.6 & 0.306 & 0.225 & \dots & \dots & \dots & \dots \\
\enddata
\label{photerror}
\end{deluxetable}


\begin{deluxetable}{lccc}
\tablecaption{Possible Systematic Uncertainties (in mags)}
\tablewidth{0pt}
\tablehead{
&\colhead{Photometric} &\colhead{Filter} \\
\colhead{Galaxy} &\colhead{Zero Point} &\colhead{Response} 
&\colhead{Extinction$^{a,b}$} }
\startdata
IC~342   &0.015  &0.010  &0.160 \\
M74      &0.028  &0.010  &0.038 \\
M83      &0.015  &0.010  &0.038 \\
M94      &0.040  &0.030  &0.010 \\
NGC~5068 &0.010  &0.010  &0.056 \\
NGC~6946 &0.037  &0.010  &0.191 \\
\enddata 
\label{tabUnc}
\tablerefs{a: Schlegel \etal\ 1998; b: Cardelli \etal\ 1989}
\end{deluxetable}


\begin{deluxetable}{lccll}
\tabletypesize{\scriptsize}
\tablecaption{Comparison of Distance Moduli}
\tablewidth{0pt}
\tablehead{
\colhead{Galaxy} &\colhead{$(m - M)_0$} &\colhead{$D$ (in Mpc)} 
&\colhead{Method} &\colhead{Reference}  }
\startdata
IC~342 &$27.64 \pm 0.12$ &$3.4 \pm 0.2$ &Revision of WIYN Cepheids &\citet{f+07} \\
IC~342 &$27.72 \pm 0.17$ &$3.50^{+0.28}_{-0.26}$ & PNLF & This study \\
IC~342 &27.29 & 2.87 &Blue Supergiants corrected for \ion{H}{2} regions &\citet{k+97} \\
\\
M94 & $28.21^{+0.06}_{-0.08}$ &$4.39^{+0.12}_{-0.16}$ &PNLF &This study \\
M94 & $28.42 \pm 0.18$ &$4.8 \pm 0.4$ &SBF & \citet{t+01,jensen03} \\
M94 & $28.34 \pm 0.29$ &$4.7 \pm 0.6$ &{\sl HST\/} TRGB & \citet{k+03b} \\
\\
M83 & $28.40^{+0.04}_{-0.05}$ &$4.8 \pm 0.1$  &{\sl HST\/} TRGB of group &\citet{k+07} \\
M83 & $28.43^{+0.05}_{-0.06}$ & $4.85^{+0.11}_{-0.13}$ &PNLF &This study \\
M83 & $28.32 \pm 0.13$ &$4.6 \pm 0.3$  &revised {\sl VLT\/} Cepheids &\citet{s+06} \\
M83 & $28.56 \pm 0.18$ &$5.2 \pm 0.4$  &{\sl HST\/} TRGB &\citet{k+07} \\
\\
NGC~5068 &$28.68^{+0.08}_{-0.18}$ &$5.45^{+0.20}_{-0.43}$ &Metallicity corrected PNLF &This study \\
\\
NGC~6946 &$28.85 \pm 0.15$ & $5.9 \pm 0.4$ &Brightest Blue Supergiants of group &\citet{ksh00} \\
NGC~6946 &$28.90 \pm 0.18$ & $6.0 \pm 0.5$ &Brightest Supergiants &\citet{skt97} \\
NGC~6946 &$28.92 \pm 0.21$ & $6.08^{+0.62}_{-0.56}$ & PNLF &This study \\
NGC~6946 &$29.29 ^{+0.20}_{-0.32}$ & $7.2^{+0.7}_{-1.0}$ &Expanding Photospheres Method  &\citet{schmidt92} \\
\\
M74 &$29.67^{+0.06}_{-0.07}$ & $8.59^{+0.24}_{-0.27}$ & PNLF &This study \\
M74 &$29.32 \pm 0.40$ &$7.3^{+1.5}_{-1.2}$ &Brightest Blue Supergiants &\citet{skt96} \\
M74 &$29.5^{+0.5}_{-0.7}$ &$8.1 \pm 2.3$ &Brightest Supergiants &\citet{h+05} \\
M74 &$29.9^{+0.6}_{-0.8}$ &$9.6 \pm 2.8$ &SN~II-P expansion rate &\citet{h+05} \\
\\
\enddata
\tablecomments{Galaxies are listed by increasing distance and distance moduli are listed by increasing uncertainty.}
\label{tabDMod}
\end{deluxetable} 

\end{document}